\documentclass[iop,apj]{emulateapj}

\usepackage{natbib}
\usepackage{color}
\usepackage{amsmath,empheq}
\usepackage{multirow,booktabs,threeparttablex}
\usepackage{hyperref}
\hypersetup{
  pdfinfo={
    Title={Vortex Formation and Evolution in Planet Harboring Disks under Thermal Relaxation},
    Author={A. Lobo Gomes, H. Klahr, A. L. Uribe, P. Pinilla, C. Surville},
    Subject={Astrophysics},
    Keywords={Accretion disks --- Astrophysics: planet-disk interactions, protoplanetary disks --- Hydrodynamics --- Methods: numerical --- Physics: instabilities}
  }
}

\shorttitle{Vortex Formation and Evolution in Planet Harboring Disks under Thermal Relaxation}
\shortauthors{Lobo Gomes, Klahr, Uribe, Pinilla, \& Surville}
\slugcomment{Submitted: May 06, 2015; Accepted: August 04, 2015}

\begin{document}

\title{Vortex Formation and Evolution in Planet Harboring Disks under Thermal Relaxation}

\author{Aiara Lobo Gomes$^{1,2}$, Hubert Klahr$^{1}$, Ana Lucia Uribe$^{3,4}$, Paola Pinilla$^{5}$ \& Cl\'ement Surville$^{6,1}$}

\affil{$^{1}$Max-Planck-Institut f\"ur Astronomie, K\"onigstuhl 17, D-69117, Heidelberg, Germany}
\affil{$^{2}$Member of the International Max Planck Research School for Astronomy and Cosmic Physics at the University of Heidelberg, Germany}
\affil{$^{3}$Department of Physics and Astronomy, College of Charleston, 66 George St, SC 29424, Charleston, USA}
\affil{$^{4}$University of Chicago, 5801 South Ellis Avenue, IL 60637, Chicago, USA}
\affil{$^{5}$Leiden Observatory, Leiden University, P.O. Box 9513, 2300RA, Leiden, The Netherlands}
\affil{$^{6}$Institute for Computational Science, University of Zurich, Winterthurerstrasse 190, 8057, Zurich, Switzerland}

\email[email:]{gomes@mpia.de}


\begin{abstract}

We study the evolution of planet-induced vortices in radially stratified disks, with initial conditions allowing for radial buoyancy. For this purpose
we run global two-dimensional hydrodynamical
simulations, using the PLUTO code. Planet-induced vortices are a product of the Rossby wave instability (RWI) triggered in the edges of a planetary gap. In
this work we assess the influence of radial buoyancy for the development of the vortices. We found that radial buoyancy leads to
smoother planetary gaps, which generates weaker vortices. This effect is less pronounced for locally isothermal and quasi-isothermal (very small cooling rate)
disks. We observed the formation of two generations of vortices. The first generation of vortices is formed in the outer
wall of the planetary gap. The merged primary vortex induces accretion, depleting the mass on its orbit. This process creates a surface density
enhancement beyond the primary vortex position. The second generation of vortices arise in this surface density enhancement,
indicating that the bump in this region is sufficient to trigger the RWI. The merged secondary vortex is a promising explanation for
the location of the vortex in the Oph IRS 48 system. Finally, we observed a nonmonotonic behavior for the vortex lifetimes as a function of the thermal
relaxation timescale, agreeing with previous studies. The birth times of the secondary vortices also display a nonmonotonic behavior, which
is correlated with the growth time of the primary vortex and its induced accretion.

\end{abstract}


\section{Introduction}\label{sec:intro}

	High mass planets leave remarkable features in their parent protoplanetary disks (PPDs), namely a gap, spiral waves, vortices, and
eccentricities. These
features are captured in
numerical simulations of planet-disk interactions
\citep[e.g.,][]{bib:nelson00,bib:winters03,bib:klahr06,bib:kley06,bib:valborro07,bib:uribe11,bib:lin11a,bib:lin11b,bib:ataiee13,bib:zhu14}, and are also expected to be observationally
detectable \citep{bib:regaly10,bib:pinilla12,bib:ruge13,bib:regaly14,bib:ruge14,bib:juhasz14,bib:pinilla15}. In this work we are particularly interested in studying the evolution of planet-induced vortices in
buoyantly unstable disks.

	Vortices can be formed in PPDs as a product of a Kelvin-Helmholtz instability, referred to as the Rossby wave instability (RWI) for accretion
disks, and/or unstable radial buoyancy. The RWI can be
triggered when there is a local bump in the inverse potential vorticity profile of the
disk \citep{bib:lovelace99,bib:li00}. Radial buoyancy can be manifested as the baroclinic instability \citep[BI,][]{bib:klahr03}, which needs a radially
decreasing pressure and entropy, or in other words, a pressure and
entropy gradients with the same sign. Vortices can be amplified due to the subcritical baroclinic instability \citep[SBI,][]{bib:lesur10}, which is a
nonlinear process. A convective overstability \citep[CO,][]{bib:klahr14,bib:lyra14} is also able to amplify vortices, CO is linear phase of SBI. More
about this topic will be discussed in Section \ref{sec:vortices}. Vortices are
interesting structures to be studied, considering that they are important in the context of planet formation, angular momentum transport through the dead zone, and type I migration.

	In the context of planet formation, vortices are good candidates to trap dust grains allowing them to grow to planetesimal or planets sizes
\citep{bib:barge95,bib:klahr06b}. This scenario is a possible
solution for the radial drift barrier -- large dust grains achieve high velocities toward the central star, making for them impossible to grow before
being accreted \citep{bib:whipple72}. However, if
the disk has a pressure bump, the dust grains can get trapped into this pressure maximum and an anticyclonic vortex is an example of such maxima
\citep[e.g.,][]{bib:bracco99,bib:varniere06,bib:inaba06,bib:lyra09b,bib:regaly12,bib:meheut12b}.

	Accretion disks need some mechanism to transport angular momentum
outwards, allowing then matter to fall inwards. \cite{bib:shakura73} introduced an $\alpha$-disk model to explain this transport, where viscosity,
triggered by some kind of turbulence, is shown to be an
efficient accretion mechanism. Usually, magnetorotational instability \citep[MRI;][]{bib:balbus91} is the most invoked mechanism to explain
turbulence in accretion disks, though in PPDs
there is a region, around the disk's midplane, where the level of ionization is not high enough for MRI to take place: the so called dead zone \citep{bib:gammie96}. The problem of angular momentum transport through the dead zone has been
investigated by many authors \citep[e.g.,][]{bib:klahr03,bib:lesur10,bib:dzyurkevich10,bib:meheut12a}. Large-scale vortices in the dead zone of PPDs can
help to transport angular momentum through that region. \cite{bib:meheut12a} studied the angular
momentum flux carried by Rossby vortices. The exchange of angular momentum between Rossby waves in the inner and outer sides of a density bump,
leads to a negative net flux, thus an outward transport of angular momentum.

	Vortices may also play a role in the context of type I migration. Planet cores and low mass planets experience type I migration
\citep{bib:ward97,bib:tanaka02}. One of the biggest issues about type I
migration is the fast time scale in which it happens. Vortices are able to trap not only dust particles but also planet cores,
thus they are able to slow down the type I migration rate
\citep[e.g.,][]{bib:koller03,bib:ou07,bib:li09,bib:yu10,bib:regaly13,bib:ataiee14}.

	The formation of planet-induced vortices is being explored thoroughly
\citep[e.g.,][]{bib:balmforth01,bib:valborro07,bib:lyra09,bib:lin11b,bib:zhu14,bib:fu14,bib:les15}. \cite{bib:fu14} studied the long term evolution of vortices
depending on the disk viscosity, disk temperature, and planet mass. They found critical parameters for the disk viscosity
($\nu = 10^{-7}r_p^2\Omega_p^2$) and temperature ($h/r_p = 0.06$) that lead to a long vortex lifetime ($\sim1$~Myr). A nonmonotonic behavior with
respect to the viscosity and temperature was found, thus high and low viscosities/temperatures lead to a faster damping of the vortices. They
concluded also that disks with same viscosity and temperature, but more massive planets, in their case $5M_J$,
can sustain vortices for a longer time. \cite{bib:les15} studied vortex evolution in terms of different cooling timescales. They found a non-monotonic
dependence of the vortex lifetimes with the cooling timescales, which is in agreement with \cite{bib:fu14}. Moreover, they pointed out the importance
of not considering locally isothermal disks, due to the fact that the RWI theory was developed for adiabatic disks \citep{bib:lovelace99,bib:li00}.

	In addition to the theoretical/numerical stage of this field, observations with high angular resolution are increasing. The {\it Atacama Large
Millimeter/submillimiter Array}
(ALMA) is now giving the capabilities to detect structures which may be related with unseen planets. Recently, dust
asymmetries were observed in five different systems: LkHa 330 \citep{bib:isella13}, Oph IRS 48 \citep{bib:marel13}, HD 142527
\citep{bib:casassus13,bib:fukagawa13}, SAO 206462 \citep{bib:perez14}, and SR 21 \citep{bib:perez14}. An anticyclonic vortex could be a reasonable
explanation for these asymmetries; however, the definite explanation for these observations is still under debate \citep{bib:pinilla15,bib:flock15}.

	The aim of this work is to study the long term evolution of planet-induced vortices in buoyantly unstable disks. The paper is laid out as follows.
In Section \ref{sec:simulations} we describe the planet-disk model and simulation setups. We describe the general evolution of our different
simulations in Section \ref{sec:results}. We discuss the role that the RWI and buoyancy played for vortex formation and sustenance in Section \ref{sec:vortices}. 
We study the convergence of our results with respect to several factors in Section \ref{sec:code_check}. We observed the formation of a second
generation of vortices, which arise in a surface density enhancement that is created beyond the primary vortex position. The formation of the
secondary vortices is discussed in Section \ref{sec:2ndvortex}. The vortex lifetimes and birth times with respect to different thermal relaxation
timescales is discussed in Section \ref{sec:lifetime}. Lastly, in Section \ref{sec:conclusions} we
briefly summarize our results and state our conclusions.


\section{Simulations}\label{sec:simulations}

	We study the formation and evolution of vortices in the outer edge of planetary gaps by solving numerically the following system of hydrodynamical (HD) equations
\begin{empheq}[]{align}
\frac{\partial \Sigma}{\partial t} + \boldsymbol{\nabla} \cdot (\Sigma \mathbf{v}) = 0, \label{eq:HDsigma} & \\
\frac{\partial \mathbf{v}}{\partial t} + \mathbf{v} \cdot \boldsymbol{\nabla} \mathbf{v} = -\frac{\boldsymbol{\nabla} p}{\Sigma} -\boldsymbol{\nabla} \Phi _{g}, \label{eq:HDv} & \\
\frac{\partial p}{\partial t} + \mathbf{v} \cdot \boldsymbol{\nabla} p + \Sigma c_s^2 \boldsymbol{\nabla} \cdot \mathbf{v} = 0, \label{eq:HDp} &
\end{empheq}
where $\Sigma$ is the gas surface density, $\mathbf{v}$ is the velocity, $p$ is the vertically integrated
pressure, $\Phi_{g}$ is the gravitational potential, and $c_{s}$ is the sound speed. In order to close the system of equations, we used an
ideal equation of state $p = c_s^2 \Sigma/\gamma$, with $\gamma = 1.4$. 

	We considered an inviscid disk, thus no prescribed viscosity was included.
This approximation may influence the vortex evolution, since previous works showed that the vortex lifetime is inversely proportional to the magnitude of
viscosity \citep{bib:valborro07,bib:ataiee13,bib:fu14}. In this work, we would like to study the direct influence of the RWI and radial buoyancy for
the development of the vortices. Therefore we chose to not include a prescribed viscosity. In our models, the only possible source of viscosity is the
turbulence-triggered viscosity by the hydrodynamical instabilities. Lastly, we assumed that the barycenter of the system is located
at the star's center. This simplification is plausible, because the planet masses considered are not very large ($1$ and $3M_J$) neither the
vortices accumulate much mass,\footnote[7]{We obtained vortices masses up to a few $10^{-4}M_{\odot}$, integrating the surface density with respect to the area element
inside the vortex region.} thus the deviation of the barycenter
with respect to the star's center should be small. Nonetheless, this approximation may slightly influence planet-induced vortex formation, since it
eliminates the Lagrange point L3, in the corotation region, which could change the gap structure.

	We used the planet-disk module for the PLUTO code that is presented in \citet{bib:uribe11}. The gravitational potential includes contributions from the planet and
the star, and it is given by
\begin{equation}
\Phi _g (\mathbf{r}) = - \frac{GM_p}{\sqrt{(\mid \mathbf{r} - \mathbf{R_p} \mid ^2 + \epsilon ^2)}} - \frac{GM_{\star}}{\mid \mathbf{r} \mid},
\end{equation}
where $G$ is the gravitational constant, $M_p$ is the planet mass, $\mathbf{R_p}$ is the planet location, $M_{\star}$ is the stellar mass and $\epsilon$ is a softening parameter. It is
needed to soften the gravitational potential of the planet in order to avoid numerical divergence close to the planet's location. Moreover, this
softening can account for 3D effects of vertical stratification. We considered this parameter as being a fraction
of the Hill radius $\epsilon ~=~kR_H$, with $k=0.6$ and
$R_H ~=~R_p(M_p/(3M_{\star}))^{1/3}$. The recommended softening factor for the planet gravitational potential is of $\epsilon = 0.6H - 0.7H$
\citep{bib:kley12}, where $H$ is the disk scale height. These values can recover 3D effects of vertical stratification. The Hill radius and the disk scale height at the planet position are
similar in our simulations, thus we chose $\epsilon = 0.6R_H$.

	The stationary solution of a sub-Keplerian disk was taken as initial conditions, which in polar coordinates is given by
\begin{empheq}[]{align}
& \Sigma = \Sigma_0 \Big ( \frac{r}{r_0}\Big )^{-\beta_{\Sigma}}, \\
& c_s = c_0 \Big (\frac{r}{r_0}\Big )^{-\beta_T/2}, \\
& v_r = 0, \\
& v_{\phi} = \sqrt{v_K^2 + \frac{r}{\Sigma}\frac{\partial p}{\partial r}},
\end{empheq}
where $\Sigma_0$ is the initial surface density at $r_0 = 1$~AU, $\beta_{\Sigma}~=~1.5$ is the slope
for the power law distribution of surface density, $\beta_T/2 = 0.5$ is the slope for the power
law distribution of sound speed, $v_K$ is the Keplerian velocity, and $h = c_s/v_K = H/r = 0.05$ is the initial aspect ratio and fix the intial sound
speed $c_0$ at $r_0 = 1$~AU, since $v_K$ at $r_0 = 1$~AU is set as one.

	The planet is set up as a point mass in a given position $R_p$ and with
a given mass $M_p$. In order to avoid an initial big disturbance to the disk, we added
the planet slowly along its first Keplerian orbit, according to the following
\begin{equation}
M_p' = M_p \Big [ \sin \Big ( \frac{\pi}{2}\frac{t}{P} \Big ) \Big ] ^2,
\label{eq:mp}
\end{equation}
where $t$ is the global time and $P = 2\pi(GM_{\star}/R_p)^{-3/2}$ is one planetary
orbit. Thus, while $t<T$ the planet mass slowly increases toward $M_p$.

	The planet initial velocity is assumed to be the Keplerian velocity $v_{\phi,p} =
\sqrt{GM_{\star}/R_p}$ and the initial acceleration coming from the gravitational
interaction with the star and the disk is
\begin{equation}
\mathbf{a_p} = -\frac{GM_{\star}\mathbf{R_{p}}}{\mid \mathbf{R_{p}}
\mid ^3} + \xi\int_{\text{disk}} \frac{G\Sigma(\mathbf{r}
- \mathbf{R_{p}}) }{\sqrt{(\mid \mathbf{r} - \mathbf{R_{p}} \mid^2 + \epsilon^2 )^3}}dA,
\label{eq:ap}
\end{equation}
where $dA$ is the area element and $\xi$ is a factor that soften the
contribution of the disk gravity in the Hill sphere and is given by
\begin{equation}
\xi = 1 - \exp \Big [ - \frac{\mid \mathbf{r} - \mathbf{R_{p}} \mid
^2}{(0.6 R_H)^2} \Big ].
\end{equation}

	Lastly, the planet position, velocity, and acceleration changed
according to the dynamical interaction with the star-disk system. Its
acceleration changes with time following Equation \ref{eq:ap} and its position and velocity are then updated using a leapfrog integrator.

\subsection{Thermal Relaxation}

	In order to account for radiative effects, we applied cooling to the system. We modeled this cooling via thermal relaxation, following the approach below
\begin{equation}
\frac{dT}{dt} = -\frac{(T - T^0)}{\tau(r)},
\label{eq:TR}
\end{equation}
where $T$ is the temperature, $T^0$ is the initial temperature (equilibrium temperature as result of irradiation), and $\tau(r)$ is the relaxation timescale, which depends on radius
($\tau(r) = 2\pi\tau/\Omega(r)$). This approach tends to reestablish
the equilibrium temperature profile, after the planetary gap is opened and the system reaches a steady state.

	Instead of adding cooling as a source term in the energy equation, we updated the temperature at each time step according to equation \ref{eq:TR}. Numerically it corresponds to
\begin{equation}
T^{\text{new}} = T^{\text{old}} - \frac{\Delta t}{\tau(r)}(T^{\text{old}} - T^0),
\label{eq:TR2}
\end{equation}
where $T^{\text{new}}$ is the relaxed temperature, $T^{\text{old}}$ would be the temperature we get from the solution of the energy equation, and $\Delta t$
is the time step. We solve Equation \ref{eq:HDp}, which describes conservation of energy,
considering pressure as an independent variable. Hence, we had to convert equation \ref{eq:TR2} from temperature to pressure dependent, for which we used the
relation $T \propto p/\Sigma$, leading to
\begin{equation}
p^{\text{new}} = p^{\text{old}} - \frac{\Delta t}{\tau(r)}\Big (p^{\text{old}} - p^0\frac{\Sigma^{\text{new}}}{\Sigma^0}\Big ),
\label{eq:PR}
\end{equation}
where $p^{\text{new}}$ is the new pressure from the relaxed temperature, $p^{\text{old}}$ is the pressure we get from the solution of equation \ref{eq:HDp}, $p^0$ is the initial pressure,
$\Sigma^{\text{new}}$ is the density we get from the solution of equation \ref{eq:HDsigma} and $\Sigma^0$ is the initial density. Finally, we cooled
the disk through equation \ref{eq:PR}. For the locally isothermal setups, instead of using Equation \ref{eq:PR} to cool the disk, we setup the
sound speed to its initial profile at every time step, in order to guarantee locally isothermality.

\subsection{Numerical Setup}

	The simulations were carried out using the finite volume Godunov-type code
PLUTO \citep{bib:pluto}. Spacial integration and time evolution were performed
using the piecewise parabolic method (PPM) and second order Runge-Kutta schemes, respectively. The Harten-Lax-van Leer-Contact (HLLC) Riemann solver was used to compute the numerical fluxes and the Strang operator splitting method to solve
the equations in multi-dimensions.

	The HD equations were solved in a two-dimensional domain considering polar coordinates. A logarithmic grid was used for the radius and a uniform one for azimuth. The system was integrated from 0.25~AU to 4.0~AU in radius and from 0 to $2\pi$ in azimuth. Temporal evolution was taken up to
5000~orbits. Reflective boundary conditions were used in the radial direction and
periodic conditions in the azimuthal direction. Distances are given in units of $1$~AU; surface densities in units of $\Sigma_0 =
10^{-4}~M_{\odot}/\text{AU}^{2}$, which corresponds to a
disk mass of $0.002M_{\odot}$ inside the domain considered, therefore it is plausible to neglect disk self-gravity, since the Toomre parameter is
$Q >> 1$ everywhere in the disk;
and velocities in units of Keplerian speed at $1$~AU. Table
\ref{tab:parameters} summarizes the simulations parameters.

\begin{table}[!htb]\centering
\caption{Simulations parameters}
\begin{threeparttable}
\begin{tabular}{ccccc}\toprule\toprule

Label & $M_p$\tnote{a} & $R_p$\tnote{b} & $\tau$\tnote{c} & $(N_r, N_{\phi})$\tnote{d} \\
 & $(M_J)$ & (AU) & ($2\pi/\Omega_o$) & \\\midrule

TR001 & 1.0  & 1.0 & 0.01 & (512, 1024) \\  
TR01  & 1.0  & 1.0 & 0.1  & (512, 1024) \\  
TR1   & 1.0  & 1.0 & 1.0  & (512, 1024) \\  
TR2   & 1.0  & 1.0 & 2.0  & (512, 1024) \\  
TR5   & 1.0  & 1.0 & 5.0  & (512, 1024) \\  
TR10  & 1.0  & 1.0 & 10.0 & (512, 1024) \\  
ISO1MJ  & 1.0  & 1.0 & 0.0  & (512, 1024)  \\
ISO3MJ  & 3.0  & 1.0 & 0.0  & (512, 1024)  \\\bottomrule

\end{tabular}

\begin{tablenotes}
{\footnotesize
\item[a] Planet mass in terms of Jupiter mass (considering $M_{\star} = M_{\odot}$).
\item[b] Planet location in AU.
\item[c] Thermal relaxation timescale in orbital units.
\item[d] Numerical resolution in the radial ($N_r$) and azimuthal ($N_{\phi}$) directions.
}
\end{tablenotes}

\label{tab:parameters}
\end{threeparttable}
\end{table}


\section{General Evolution}\label{sec:results}

	In this section we describe the system evolution for our simulations. First, we present the results for the simulations with a $1M_J$ planet
mass, varying the thermal relaxation timescales. Second, we present the results for the isothermal simulations, considering $1M_J$ and $3M_J$
planet masses.

\subsection{Non-isothermal Cases}

	The first set of results presented are the cases with a $1M_J$ planet and different thermal relaxation timescales. The
different values of $\Omega\tau$\footnote[8]{Hereafter, we refer to $\tau(r)$ as $\Omega\tau$.} considered and their labels can be seen in Table
\ref{tab:parameters}. All the simulations presented a similar behavior, which we describe hereafter.

	The formation of spiral waves takes place during the first planetary orbit.
Additionally, during the first tens of orbits the planet carves out a very noticeable gap and small vortices are formed in the outer edge of this gap. In
the first hundreds of planetary orbits these small
vortices merge into a bigger one. Some mass remains in the gap region even after a few thousands of orbits, which can be related to the inviscid
and/or non-barycentric approximations. The inviscid approximation may influence the efficiency of mass transport. Nonetheless, neglecting the
indirect potential exerted on the disk due to the barycenter shift also seems to retain mass in the gap region, even for non-inviscid disks \citep[see Figure 2 in][]{bib:zhu14}. In the first thousands of planetary orbits a surface density enhancement appears beyond the vortex position. Accumulation
of mass persists and a second vortex is formed in this region. The primary vortex is damped in different
timescales for the different $\Omega\tau$'s, nonetheless some material remains in a ring-shape
form in between the planetary gap and the secondary vortex. This material is also dispersed out in different timescales.

\begin{figure*}[!htb]\centering
\includegraphics[width=.91\textwidth]{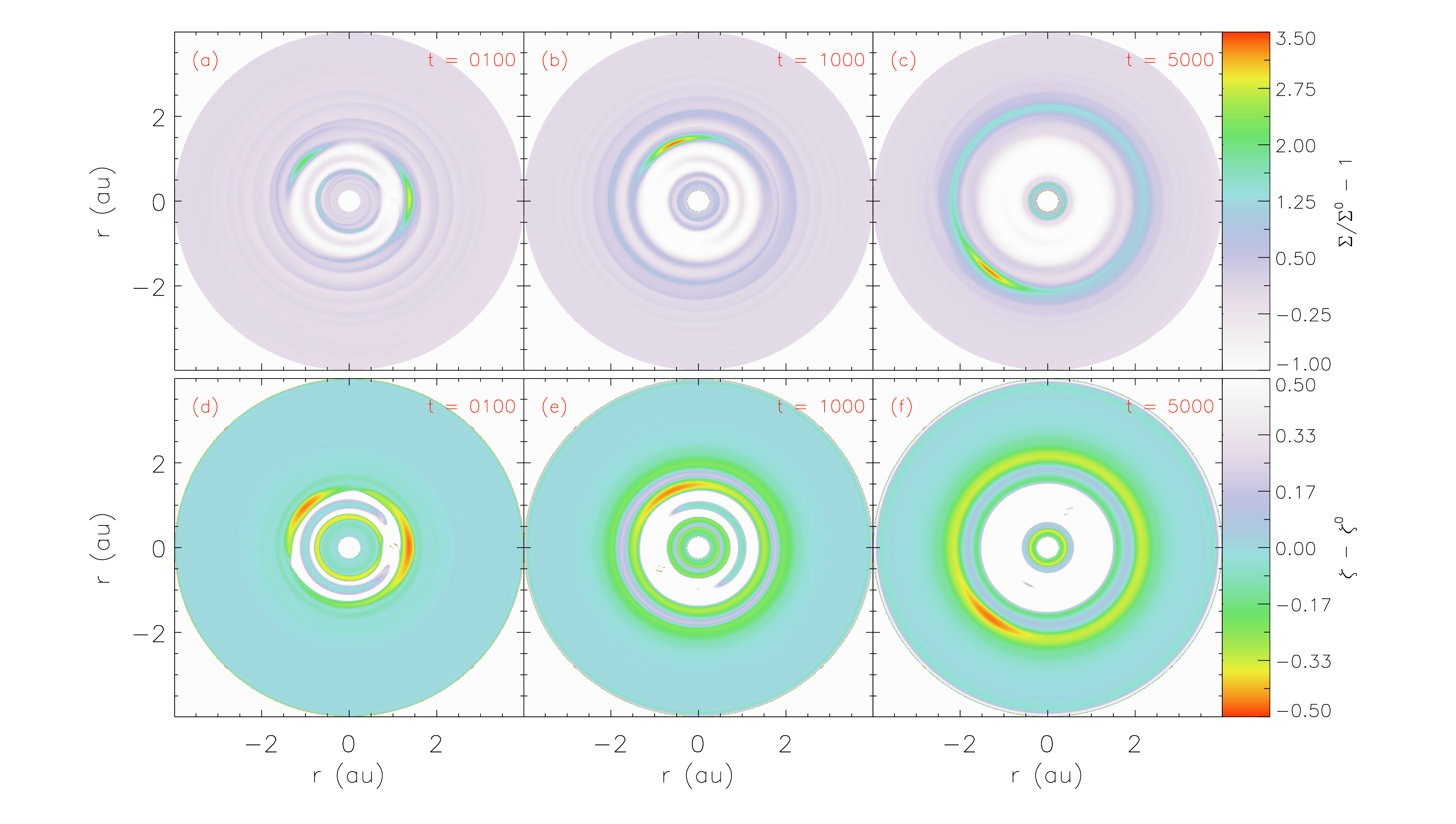}
\caption{Evolution of the surface density perturbation (top panel) and the potential vorticity with the Keplerian profile subtracted (bottom panel).
The color bar for the potential vorticity plots was truncated from $-0.5$ to $0.5$, in order to provide a higher contrast. The results show simulation TR01.}
\label{fig:TR01}
\end{figure*}

\begin{figure*}[!htb]\centering
\includegraphics[width=.91\textwidth]{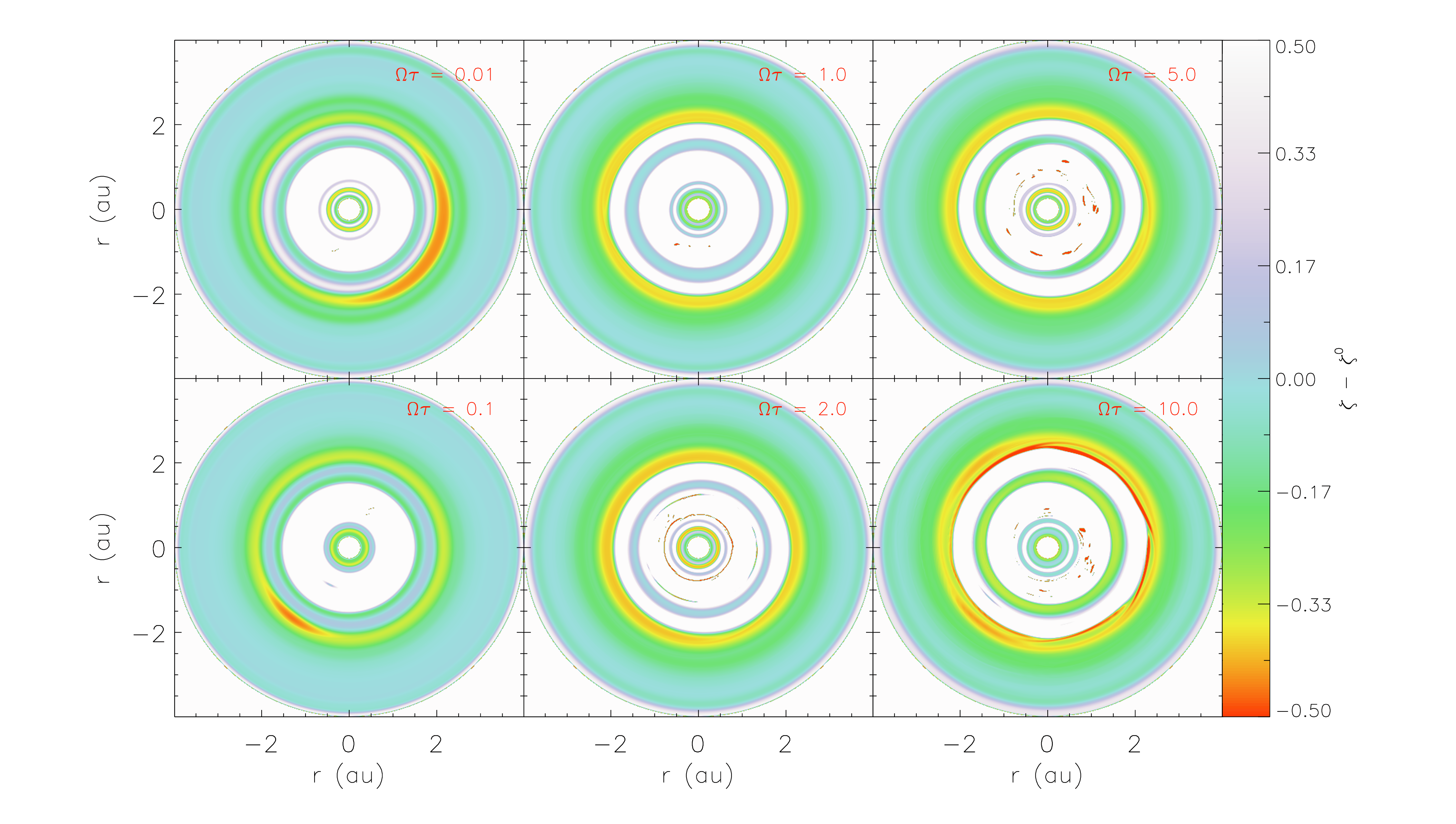}
\caption{Final potential vorticity with the Keplerian profile subtracted for the different $\Omega\tau$'s considered.}
\label{fig:TRruns}
\end{figure*}

	Simulations TR001 and TR01 present a secondary vortex very similar to the primary one. Figure \ref{fig:TR01} shows the system evolution
for TR01. Simulations TR1, TR2, TR5, and TR10 present also a secondary vortex; however, the new vortex is highly spread in the azimuthal direction. Figure \ref{fig:TRruns} shows the potential vorticity at $5000$~orbits for the
different $\Omega\tau$'s. For $\Omega\tau~=~0.01$, the secondary vortex survives until the end of the simulation; however, the vortex gets spread
along the azimuthal direction. For $\Omega\tau~=~0.1$, the secondary vortex survives and does not get spread in the azimuthal direction. For
$\Omega\tau~\ge~1.0$, the secondary vortex is mostly damped by the end of the simulation interval.

\subsection{Isothermal Cases}

	Here we present the results for the isothermal setup and planet masses of $1M_J$ and $3M_J$. The simulations labels are presented on Table
\ref{tab:parameters}.

	The isothermal configuration shows a considerably similar behavior as the models with thermal relaxation. For the ISO1MJ simulation, the
sequence of events is the same. We first observe
the formation of spiral waves, followed by planet gap opening, and production of small
vortices at the outer edge of this gap. The small vortices gather together and merge into a bigger one. A surface density enhancement appears beyond
the primary vortex position. Material is accumulated at this location and a second vortex arises, this structure gets spread in the azimuthal direction with
time. The primary vortex gets damped and the material in between the planetary gap and the secondary vortex disperses out. The timescales for the events are similar to the ones
for the non-isothermal cases. 

\begin{figure*}[!htb]\centering
\includegraphics[width=.91\textwidth]{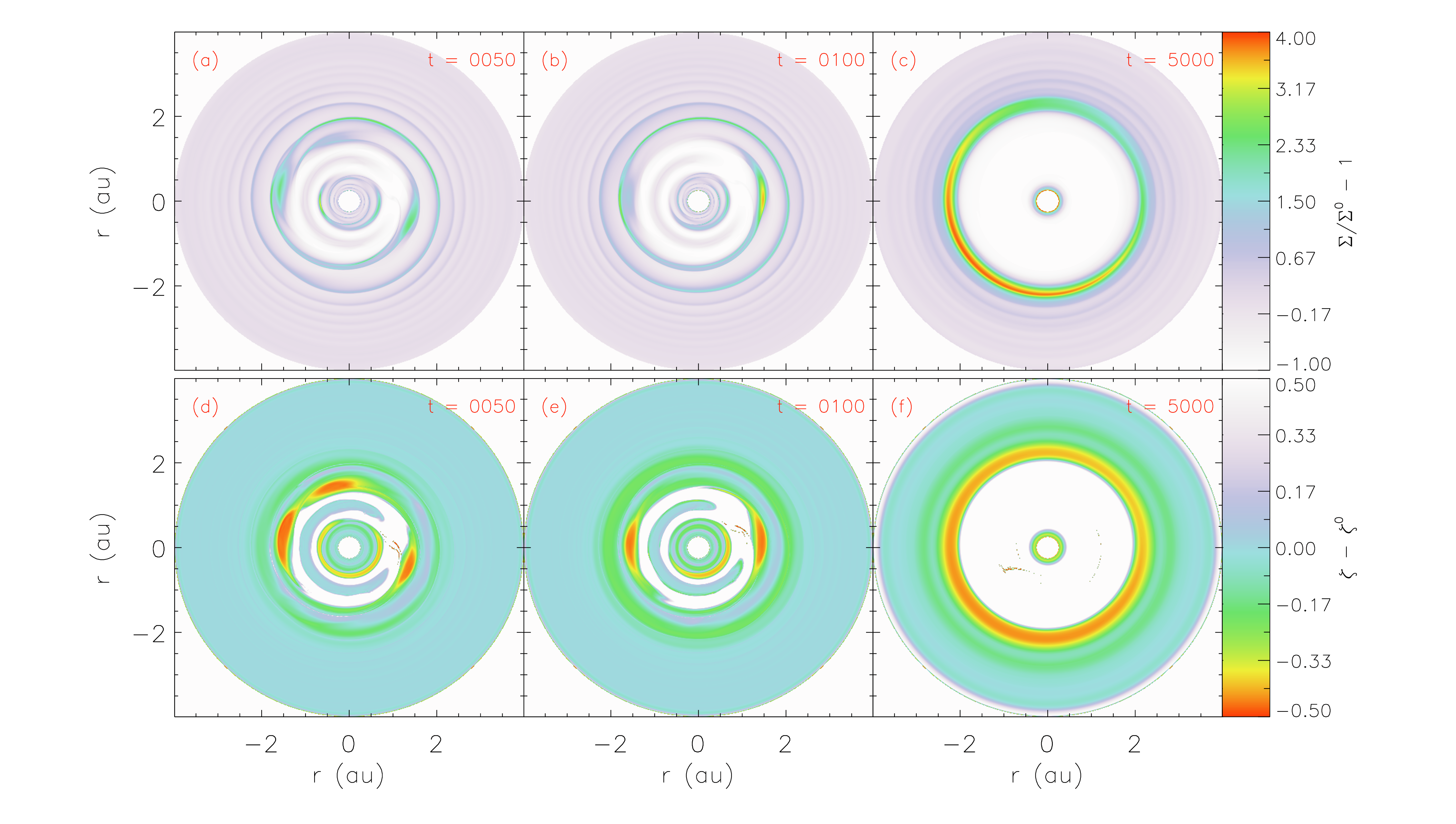}
\caption{Evolution of the surface density perturbation (top panel) and the potential vorticity with the Keplerian profile subtracted (bottom panel).
The color bar for the potential vorticity plots was truncated from $-0.5$ to $0.5$, in order to provide a higher contrast. The results shows
simulation ISO3MJ.}
\label{fig:ISO3MJ}
\end{figure*}

	The ISO3MJ simulation presents a similar sequence of events, with the difference that two vortices, that do not merge with time, are created
in the outer edge of the planetary gap. The
evolutionary timescales for which different structures form are also different. The surface density enhancement appears in hundreds of planetary
orbits, instead of thousands. The damping of the primary vortices is also faster. Pile-up of material at the surface density enhancement also
happens.
Nonetheless, it takes thousands of planetary orbits for a secondary vortex to arise. After a few thousands of planetary orbits the material between
the planetary gap and the secondary vortex is totally dispersed out, and a much wider gap is settled. Figure \ref{fig:ISO3MJ} shows the system evolution for simulation ISO3MJ.

	It was not possible to consider a higher planet mass (e.g., $10M_J$) under the setup assumed, because the gap created is much wider, which
makes the disk size considered too small. To solve the same structures in a bigger disk, we would need to use more grid cells.


\section{Vortex Formation and Evolution}\label{sec:vortices}

	In a protoplanetary disk, we know that vortex formation can happen as a product of the RWI
\citep{bib:lovelace99,bib:li00} and/or radial buoyancy \citep{bib:klahr03,bib:lesur10,bib:klahr14}. In this work, we considered initially buoyantly unstable disks. Nonetheless, we know that the presence of a planetary gap
naturally triggers the RWI, due to the sharp surface density gradient that is created in the gap edges.
Hereafter, we discuss the role that the RWI and radial buoyancy played for the formation and evolution of planet-induced vortices. We would like to
remember that we are using an inviscid approximation, thus any viscosity in the system is turbulence-triggered viscosity by the hydrodynamical
instabilities. We chose to consider an inviscid approximation to assess the direct influence of radial buoyancy and the RWI for the vortices
evolution.

\vfill

\subsection{Rossby Wave Instability}

	The RWI is a pressure driven instability for rotating systems, which is composed of non-axisymmetric modes. The RWI is triggered when there is a
local maximum in the radial profile of the function \citep{bib:lovelace99}
\begin{equation}
\mathcal{L} (r) \equiv \mathcal{F} (r) S^{2 / \gamma} (r),
\label{eq:L}
\end{equation}
where $\mathcal{F}^{-1} = (\vec{\nabla} \times \vec{v})\cdot\hat{z}/\Sigma$ is the potential vorticity\footnote[9]{Hereafter, called as $\zeta$ instead of
$\mathcal{F}^{-1}$.} and $S = p/\Sigma^{\gamma}$ is an equivalent to the entropy. Physically, this
condition can be achieved at the edge of planetary gaps \citep{bib:valborro07} and the edge of dead zones due to sharp viscosity transitions \citep{bib:lyra12}.

	The formation of vortices as a product of the RWI has been studied by many authors. The growth rate of the instability for 2D disks was studied
by \cite{bib:li00} and \cite{bib:umurhan10}, using different approximations, and in 3D stratified disks by \cite{bib:meheut12a} and \cite{bib:lin12}.
The nonlinear phase of the instability was explored by \cite{bib:meheut13}. Despite the theory for the RWI was developed for adiabatic disks
\citep{bib:li00},
most of the works used locally isothermal disks to study the formation and evolution of planet-induced vortices
\citep{bib:balmforth01,bib:valborro07,bib:lyra09,bib:lin11b,bib:zhu14b,bib:fu14}. Just recently, \cite{bib:les15} made a breakthrough and
added an artificial source of cooling and heating to explore the non-isothermal behavior. In this context, our work is a second step to the process of
understanding the non-isothermal scenario.

\begin{figure}[!htb]\centering
\includegraphics[width=.5\textwidth]{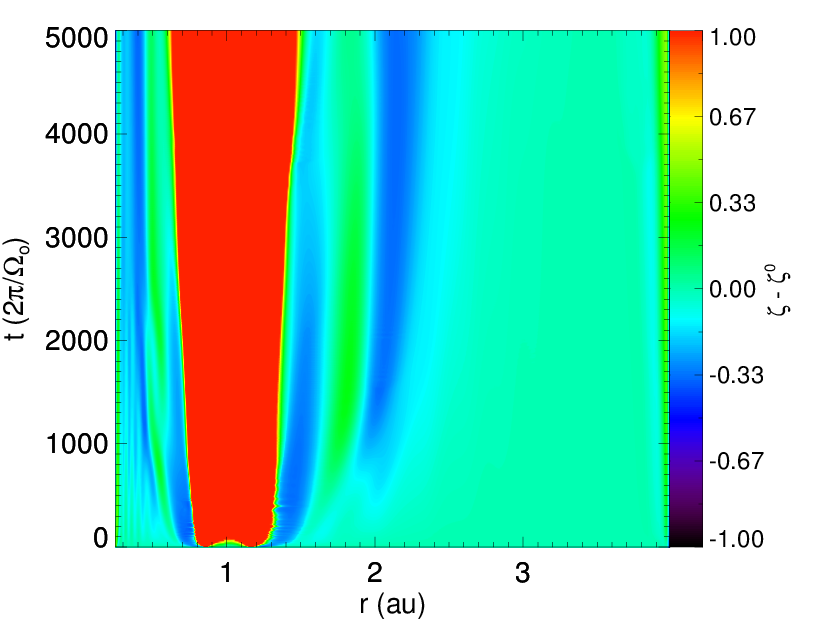}
\caption{Spacetime evolution of the potential vorticity ($\zeta$) averaged in azimuth for simulation TR01. The Keplerian profile was subtracted. The color bar was truncated from $-1.0$ to $1.0$ in order to obtain a higher contrast.}
\label{fig:zetast}
\end{figure}

	It is not in the scope of this work to make an extensive
study of the growth and decay of the RWI, since this matter was already addressed by \cite{bib:les15} for a physical scenario very similar to ours. In
order to
have just a qualitative insight, we analyzed the spacetime evolution of the potential vorticity ($\zeta$) averaged in azimuth. Figure \ref{fig:zetast} shows the result for simulation TR01.

	The blue region in Figure \ref{fig:zetast} represents a minimum in $\zeta$. A minimum in $\zeta$ is equivalent to a maximum in
$\mathcal{L}$ (Equation \ref{eq:L}), sufficient condition to trigger the RWI. This minimum is achieved in both, the outer edge of the
planetary gap and at the surface density enhancement outwards the primary vortex position. Therefore, the RWI has been triggered in both regions. The presence of the minimum is maintained along
the whole simulation interval. In the planetary gap edge, its value slowly increases with time, which explains the vortex decay. In the surface
density enhancement outwards the primary vortex position, its value is kept slightly constant with time, which explains the survival of the 
secondary vortex until the end of our simulation. The spacetime evolution of $\zeta$ is similar for all the cases. A local minimum is observed
in the regions of the primary and secondary vortices. The main differences are the size of the blue regions (local minimums), the time the local
minimum related the primary vortex starts to decay, and the time the local minimum related to the secondary vortex appears.

\subsection{Radial Buoyancy}

	The necessary ingredients for a CO and SBI are: (i) radial
pressure and entropy gradients possessing the same sign (radial buoyancy) and (ii) thermal relaxation, with maximum amplification for
$\Omega\tau \simeq 1.0$. The formation of vortices due to the BI was first observed by \cite{bib:klahr03}. Further studies by
\cite{bib:petersen07a,bib:petersen07b} showed the importance of thermal relaxation for baroclinic vortex amplification. They found that thermal
relaxation or diffusion, besides entropy gradient, are required to keep the instability in action. \cite{bib:lesur10} studied baroclinic vortex
amplification through the growth of existing vortical perturbations. In order to not cause confusion between the generation of vortices by the classical
BI and amplification of the vortices in a radial buoyant fluid, they coined this process a SBI. A parametric study covering the
important ranges of entropy gradients, thermal diffusion timescales, and thermal relaxation timescales for PPDs was carried out by \cite{bib:raettig13}.
They showed the importance of baroclinic effects even for small entropy gradients, which is the case in PPDs. \cite{bib:klahr14} found a linear
amplification of epicyclic oscillations in radially stratified and vertically unstratified disks, which they called convective overstability. This
phenomenon can be regarded as the linear phase of the SBI. Yet not much efforts were made to study how
vortex formation and amplification proceeds in a buoyantly unstable disk with a high mass planet embedded.
\cite{bib:les15} discussed briefly whether an axisymmetric instability was at play in their simulations
of planet induced vortices; however, they concluded that only the RWI was in action.
	
	We can quantify the radial stability in a disk with regards to convection through the Brunt-V\"ais\"al\"a frequency ($N$), which is given by
\citep{bib:raettig13}
\begin{equation}
N^2 = -\beta_p \beta_S \frac{1}{\gamma}\Big ( \frac{H}{r} \Big )^2 \Omega^2,
\label{eq:N}
\end{equation}
where $\beta_p$ is the pressure gradient, $\beta_S$ is the entropy gradient, and $\Omega$ is the angular velocity. Positive values of $N^2$ indicate stability. The entropy gradient
for a 2D vertically integrated disk is given by
\begin{equation}
\beta_S = \beta_T + (1 - \gamma)\beta_{\Sigma}, 
\label{eq:betaS}
\end{equation}
where $\beta_T$ is the temperature gradient and $\beta_{\Sigma}$ is the surface density gradient. We made the choice for the initial surface density and sound
speed gradients in a way that it gives an initial negative value for $N^2$ equals to $-0.0018$, thus favoring instability.

\begin{figure}[!htb]\centering
\includegraphics[width=.5\textwidth]{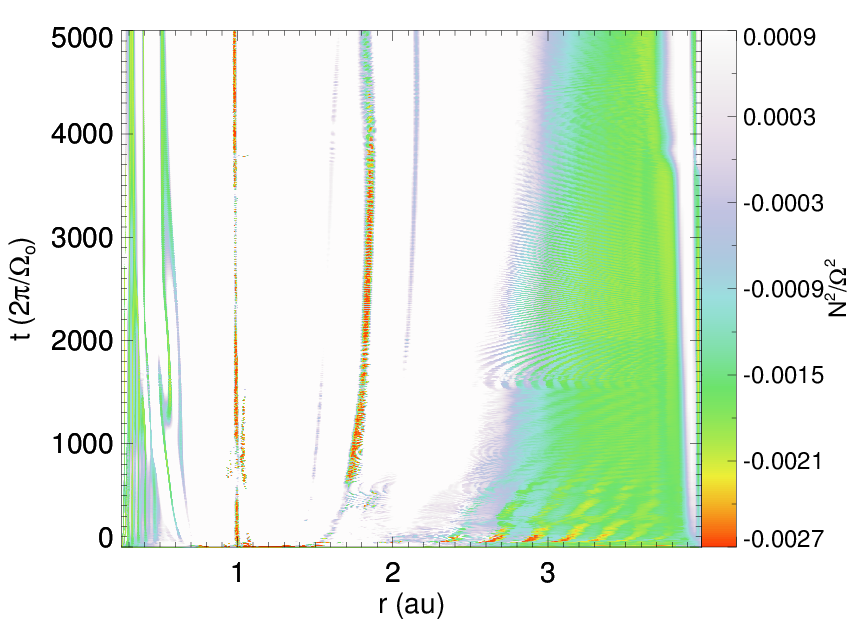}
\caption{Spacetime evolution of $N^2/\Omega^2$ averaged in azimuth for simulation TR01. The color bar was truncated from $1.5(N^2/\Omega^2)_{ini}$ to
$-0.5(N^2/\Omega^2)_{ini}$, in order to provide a higher contrast.}
\label{fig:N2st}
\end{figure}

	Figure \ref{fig:N2st} shows the spacetime evolution of $N^2/\Omega^2$ averaged in azimuth for simulation TR01. We plot $N^2/\Omega^2$
instead of $N^2$, to eliminate the dependence with the angular velocity. Since $\Omega^2$ is always positive, $N^2/\Omega^2>0$ still
indicates stability. We can see that in the outer disk $N^2$ is kept negative and roughly equals to its initial value, with exception for the outer
boundary. The outer radial extent with negative
$N^2$ becomes narrower throughout the simulation interval, thus the evolution of the system tends to stabilize the disk with respect to buoyancy. In the gap region and outer
gap wall, $N^2$ becomes positive after a few tens of planetary orbits; however, there is a strip around the primary vortex
position with smaller values of $N^2$. The strip's center possesses negative $N^2$ in the first tens of planetary orbits, but $N^2$ turns positive later on. From $\sim600$~orbits, $N^2$
becomes negative again in the central position of the vortex. The region with negative $N^2$ is not as large as the
vortex size, thus buoyancy is not playing a major role for the evolution of the primary vortex. $N^2$ is positive in the region where the second
generation of vortex appears, until the time that the secondary vortex arises and a strip with
negative $N^2$ appears around the secondary vortex position. Once more, the strip width is smaller than the vortex size, indicating that 
buoyancy may not be
playing a major role in the location of the vortex. 

	The behavior of $N^2$ for the other cases is similar to the one of TR01, with exception for TR001 and the isothermal cases. For them,
$N^2$ becomes positive in the first tens of planetary orbits and remains positive along the whole simulation interval. This indicates that 
buoyancy does
not play any role for the quasi-isothermal and isothermal cases. Reinforcing the findings of \cite{bib:petersen07a,bib:petersen07b}, in the lack of
thermal relaxation or diffusion, buoyancy is not sustained. For $\Omega\tau \geq 1.0$, the strip around the position of the primary
vortex has negative $N^2$ during a larger fraction of the primary vortex lifetime. The strip around the secondary vortex is wider,
indicating that buoyancy had
more importance for the secondary vortex evolution than in the case that $\Omega\tau < 1.0$.

\begin{figure}[!htb]\centering
\includegraphics[width=.5\textwidth]{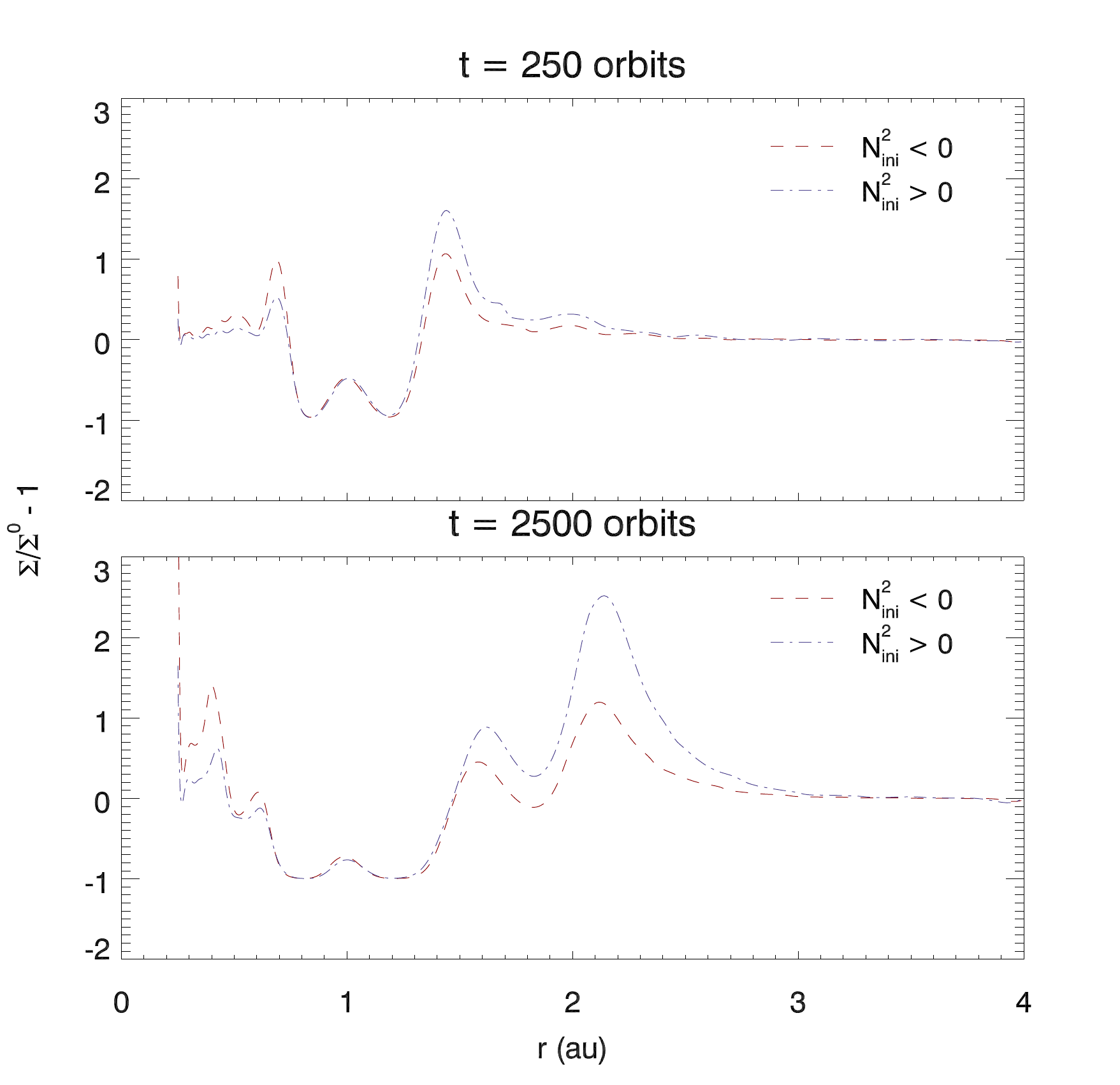}
\caption{Surface density profiles averaged in azimuth. The red dashed line shows $N^2$ initially
negative, whereas the slate
blue dotted-dashed line $N^2$ initially positive.}
\label{fig:BI}
\end{figure}

	To check the impact that buoyancy has in the results, we used a model where the initial $N^2$ is positive. We run the new simulation with the same physical and numerical setup
as for the TR01 case, but changing the surface density gradient from $\beta_{\Sigma} = 1.5$ to $\beta_{\Sigma} = 3.0$. The general evolution of the
system was very similar to the case where $N^2$ is initially negative. Two major differences were noticed. The first is regarding the maximum
amplitudes that the
primary and secondary vortices achieve, which is higher for the case where $N^2$ is initially positive. The second is regarding the second generation of
vortices. For the $N^2_{\text{ini}}>0$ case, two vortices arise in the surface density enhancement region and take more time to merge ($\sim1000$~orbits against
$\sim500$~orbits for the reference case). The secondary merged vortex has also an aspect
ratio much smaller than the secondary vortex for the $N^2_{\text{ini}}<0$ case. Figure \ref{fig:BI} presents a comparison of the surface density profiles
for the $N^2_{\text{ini}}$
positive and negative cases, for two different points in time ($t = 250$~orbits and $t = 2500$~orbits).

\begin{figure}[!htb]\centering
\includegraphics[width=.5\textwidth]{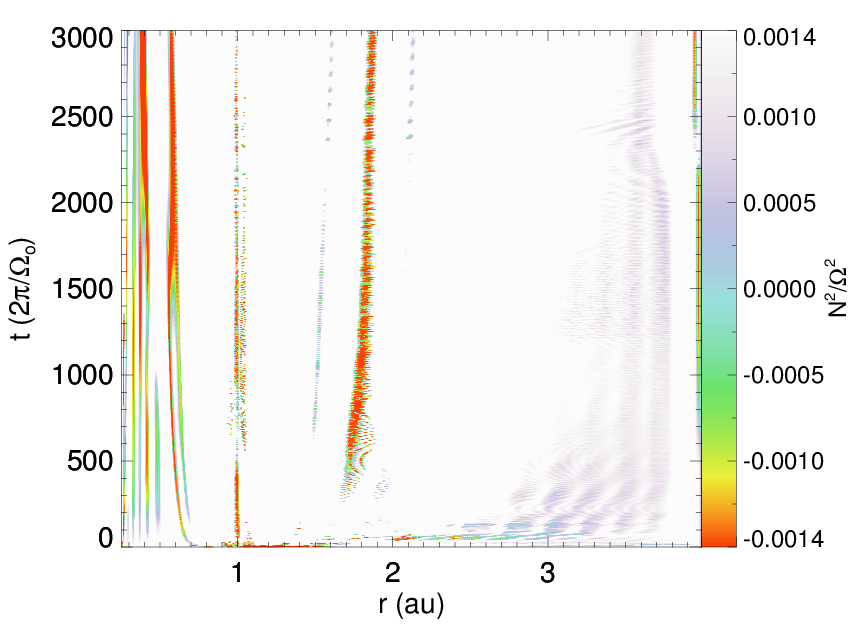}
\caption{Spacetime evolution of $N^2/\Omega^2$ averaged in azimuth for the simulation with $N^2_{ini}>0$. The color bar was truncated from $-(N^2/\Omega^2)_{ini}$ to
$(N^2/\Omega^2)_{ini}$, in order to provide a higher contrast.}
\label{fig:N2stbi}
\end{figure}

	The planetary gap structure is very similar
for the different surface density slopes, the same width and lower level for the depth are observed, as well as the same location for the maximum and
minimum surface density perturbations. The standing difference is the sharpness of the
surface density gradient in the planetary gap edge and at the surface density enhancement beyond the primary vortex position. A larger gradient produces stronger vortices, therefore the vortices for $N^2_{ini}>0$
are stronger. The times chosen to compare the cases represent a moment the vortices are totally merged. The spacetime evolution of $N^2/\Omega^2$
for $\beta_{\Sigma} = 3.0$ (Figure \ref{fig:N2stbi}) shows
that $N^2$ becomes negative in the first tens of orbits in the region where the primary vortices arise; however, it becomes positive again and
local buoyancy
disappears for hundreds of orbits, $N^2$ becomes negative again from $\sim700$~orbits. The strip with negative
$N^2$ around the vortex position is again not as wide as the vortex size, indicating that buoyancy is not playing a major role for the evolution of the
primary vortex. Nonetheless, this shows that initially
buoyantly stable disks can undergo an inversion of sign for the entropy gradient, therefore turning on instability. In the secondary
vortex region,
$N^2$ never turns to be negative. The RWI is the only responsible for the formation and sustenance of the secondary vortex.

	This result demonstrates that when we have the RWI and buoyancy
acting at the same time, weaker vortices are produced. Therefore, buoyancy opposes vortex
amplification and survival, in this scenario. Taking into account that these processes provide viscosity to the system, once we have both in action more viscosity
is produced. More viscous disks carve out smoother gaps, leading to the weaker vortices. It should also be noticed that for $\Omega\tau \geq 1.0$, the
secondary vortices get damped during the simulation interval, those are also the cases for which buoyancy plays some role for the 
secondary vortex evolution.


\section{Code Control}\label{sec:code_check}

	In this section, we explore the numerical factors that could influence the physical validity of our simulations. The tests were done using the TR01 case as reference.
The physical conditions and numerical setup were exactly the same as for TR01, varying only what we following mention.
We checked the convergence of the results considering a higher
numerical resolution. We analyzed whether our reflecting boundary conditions for the radial direction may have reflected waves, influencing the
evolution of the system. A different way to prevent boundary effects is to push the outer disk to a larger radius, hence we also used this approach to check
whether the disk size influenced the results. Lastly, we added the planet along a larger number of planetary orbits to analyse whether the initial planet disturbance could have been
too large, generating fake effects.

	For the numerical resolution test, we doubled the resolution from ($N_r$, $N_{\phi}$)~=~(512, 1024) to ($N_r$, $N_{\phi}$)~=~(1024, 2048). The temporal evolution was
taken up to 1000 planetary orbits. The full temporal evolution was not checked, because the doubled resolution is numerically highly expensive. For
the boundary conditions test, we first changed the inner and outer radial boundary conditions from reflective to non-reflective, second we considered a larger disk
extending from 0.25~AU to 8~AU. Finally, aiming to check the effect of the initial planet disturbance to the disk, we made two tests: slowly adding the
planet along its first 10 and 100 orbits, following Equation \ref{eq:mp}, in the reference case the planet was slowly added along its first
orbit. The temporal evolution for the last four tests was taken up to 5000
planetary orbits.

\begin{figure}[!htb]\centering
\includegraphics[width=.5\textwidth]{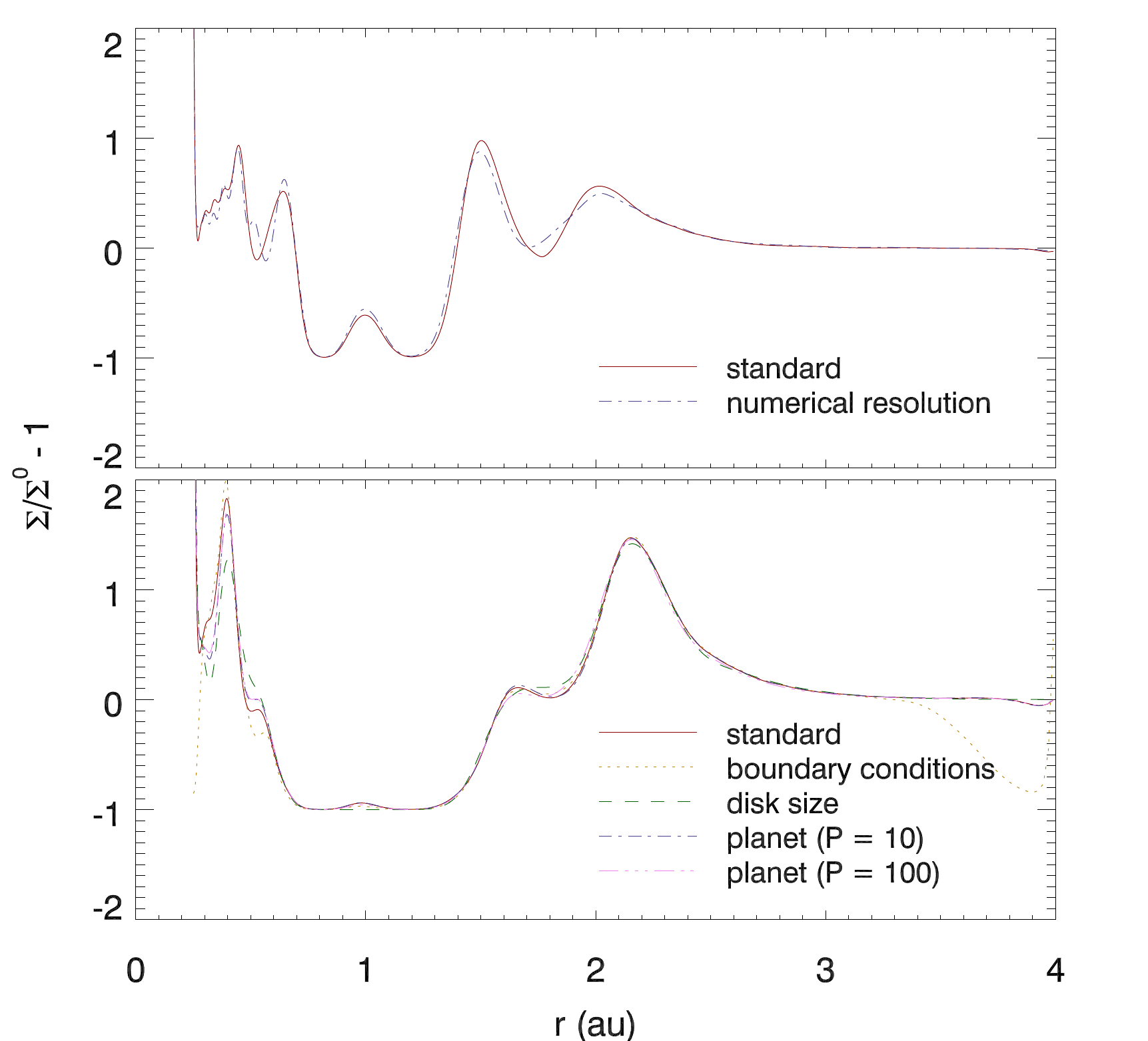}
\caption{Surface density profiles (averaged in azimuth) at $1000$~orbits (top panel) and $5000$~orbits (bottom panel). The top panel presents the comparison for the
numerical resolution test. The red solid line shows the reference case, whereas the slate blue dotted-dashed line the numerical resolution test. The bottom
panel presents the comparison for the other cases. The red solid line shows the reference case, the goldenrod dotted line shows the boundary
conditions test, the green dashed line shows the disk size test, the slate blue dotted-dashed line shows the planet disturbance test (the planet was added during its first 10
orbits), and the violet triple-dotted-dashed line shows the planet disturbance test (the planet was added during its first 100 orbits).}
\label{fig:code_check}
\end{figure}

	We compare our test cases with the reference case using their surface density profiles at the latest snapshot. Figure \ref{fig:code_check}
presents these profiles for $t=1000$~orbits (numerical resolution comparison) and $t=5000$~orbits (other comparisons). We observed a good agreement
for the surface density profiles, indicating that the simulation results were not much influenced by these factors. Nonetheless, for the resolution
test, the outer gradient that leads to the second generation of vortices is slightly smoother than for the standard case. 

	For the boundary condition test, in the outer disk ($r
\gtrsim 3$~AU) the material is emptied out, due to the boundary choice; however,
the main results agree with the standard case. The major difference is regarding the secondary vortex, for the standard case a secondary merged vortex is
created, in this case two secondary vortices are created and they do not merge until the end of the simulation. The two secondary vortices,
for the non-reflecting radial boundaries, are right opposite to each other in the azimuthal direction and they have about the same strength. We speculate that the flow in the corotation region of these vortices is not being able to push them together, exactly because the vortices have same
strength and are located right opposite to each other, leading to a stable configuration. For the larger disk
size, the standing difference is regarding the local minimum between the surface density bumps, where the primary and secondary vortices sit. For
the standard case, the local minimum is still present by the end of the simulation. For the larger disk size, this local minimum has disappeared by the
end of the simulation, meaning that the matter is dissipating slightly faster.

	For the planet being added along the first 10 orbits, the
primary small vortices merge in a faster timescale, and the vortices amplitudes take a longer time to grow than for the standard case. For the planet being added along
the first 100 orbits, the vortices amplitudes take also a longer time to grow than for the standard case. Higher planet masses can excite
stronger vortices \citep{bib:fu14}, thus when the planet perturbation is added slower, the vortices will also take a longer time to grow.

\section{Second Generation of Vortices}\label{sec:2ndvortex}

	The most interesting result of our simulations is the second generation of vortices. A surface density enhancement was observed beyond
the primary vortex position for all the cases. This bump is strong enough to trigger the RWI outside the primary vortex radius and to form a second generation of vortices. As it was
discussed before, we observe a
minimum for $\zeta$ in the region of the secondary vortex for all the cases. Also, there are strips of negative
$N^2$ in the region of the secondary vortex, with exception for the TR001 and the isothermal cases. Once more, the action of the RWI
together with buoyancy controls the vortex evolution. For the TR001 and isothermal cases, the RWI is solely responsible for the secondary vortex. We already
established that our results were not affected by the choice of boundary conditions, resolution, or the planet perturbation being too sharp. Therefore, it confirms the physical origin of the second
generation of vortices. No further density enhancement (strong enough to keep triggering the RWI) beyond the secondary vortex position was observed, even for the test simulation with a larger disk, thus we do not expect a
third generation of vortices.

\subsection{The Origin of the Secondary Vortex}

\begin{figure}[!htb]\centering
\includegraphics[width=.5\textwidth]{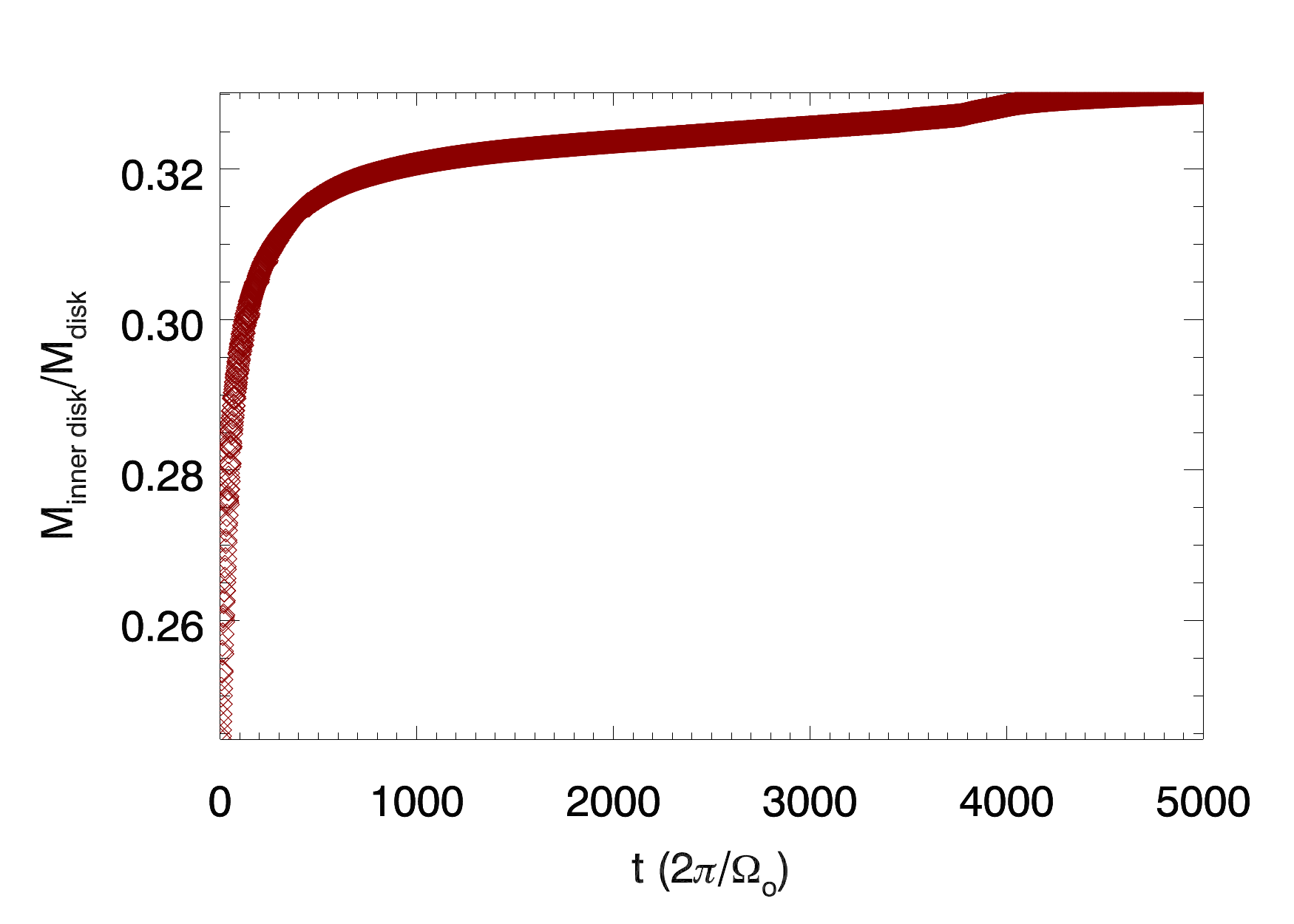}
\caption{Inner disk mass as a function of time for simulation TR01. We consider $0.25~\text{AU} \leq r \leq 0.75~\text{AU}$ as the inner disk.}
\label{fig:inmass}
\end{figure}

	Accumulation of mass is observed at the inner boundary for all the cases. Our understanding is that the primary vortices produce an effective
$\alpha$-viscosity that is large enough to induce accretion in the disk in the timescales we simulate. Therefore mass is flowing from the region of
the primary vortex to the inner disk. The depletion of mass in the orbit of the primary vortex position looks like a gap carved out by the
primary vortex. In fact,
the outer wall of the planetary gap is moving outwards due to this depletion of mass. For instance, for the ISO3MJ case, by the end of our simulation
all the mass in the orbit of the primary
vortex was already depleted and the planetary gap became wider in its outer side. Figure
\ref{fig:inmass} presents the inner disk mass as a function of time for simulation TR01, where the increase of the mass is demonstrated.

	The faster
mass increase happens in the first tens of orbits, when the planet still carving out its gap. During this period, the mass increase is mostly due to the
planetary gap opening process. The inner disk mass is being pushed from $r = 1$~AU to the inner parts of the disk. For $t \gtrsim 100$~orbits, the mass
increase is most likely due to accretion induced by the primary vortex. Our simulations assumed an inviscid disk, therefore any viscosity
is produced by the hydrodynamical instabilities, which in this case could be named a vortex-induced viscosity. We can obtain an estimate of the
$\alpha$-parameter, using the $r-\phi$ component of the Reynolds stress and the local sound speed \citep{bib:flock11}. For simulation TR01, the
$\alpha$-parameter averaged in space and time (until the time the secondary vortex appears) was $\alpha \simeq 3 \times 10^{-3}$, a value in the range of what
is typically obtained by the MRI, $\alpha = 10^{-4} - 10^{-2}$
\citep{bib:dzyurkevich10}. The other simulations presented values in the range from $\alpha = 10^{-4} - 10^{-2}$, in agreement with MRI. The
lowest value obtained was $\alpha \simeq 6 \times 10^{-4}$, for TR1, and the highest value was $\alpha \simeq 1 \times 10^{-2}$, for TR5. Large-scale
vortices are able to transport angular momentum outwards, because a negative angular momentum flux is obtained from the balance between the angular
momentum carried by Rossby waves in the inner and outer sides of a surface density bump \citep{bib:meheut12a}.

	In this work we assumed that the barycenter of the system is located at the star's center. This approximation could
influence the gap structure, since the Lagrange point L3, in the corotation region of the planet, is removed. Changes in the gap structure may
affect the primary vortex generation, subsequently possibly impacting the second generation of vortices. This assumption summed to the inviscid disk approach are
two factors that could influence the formation of a second generation of vortices. We suggest that further studies should check these factors.

\subsection{Pressure Bumps}

\begin{figure}[!htb]\centering
\includegraphics[width=.5\textwidth]{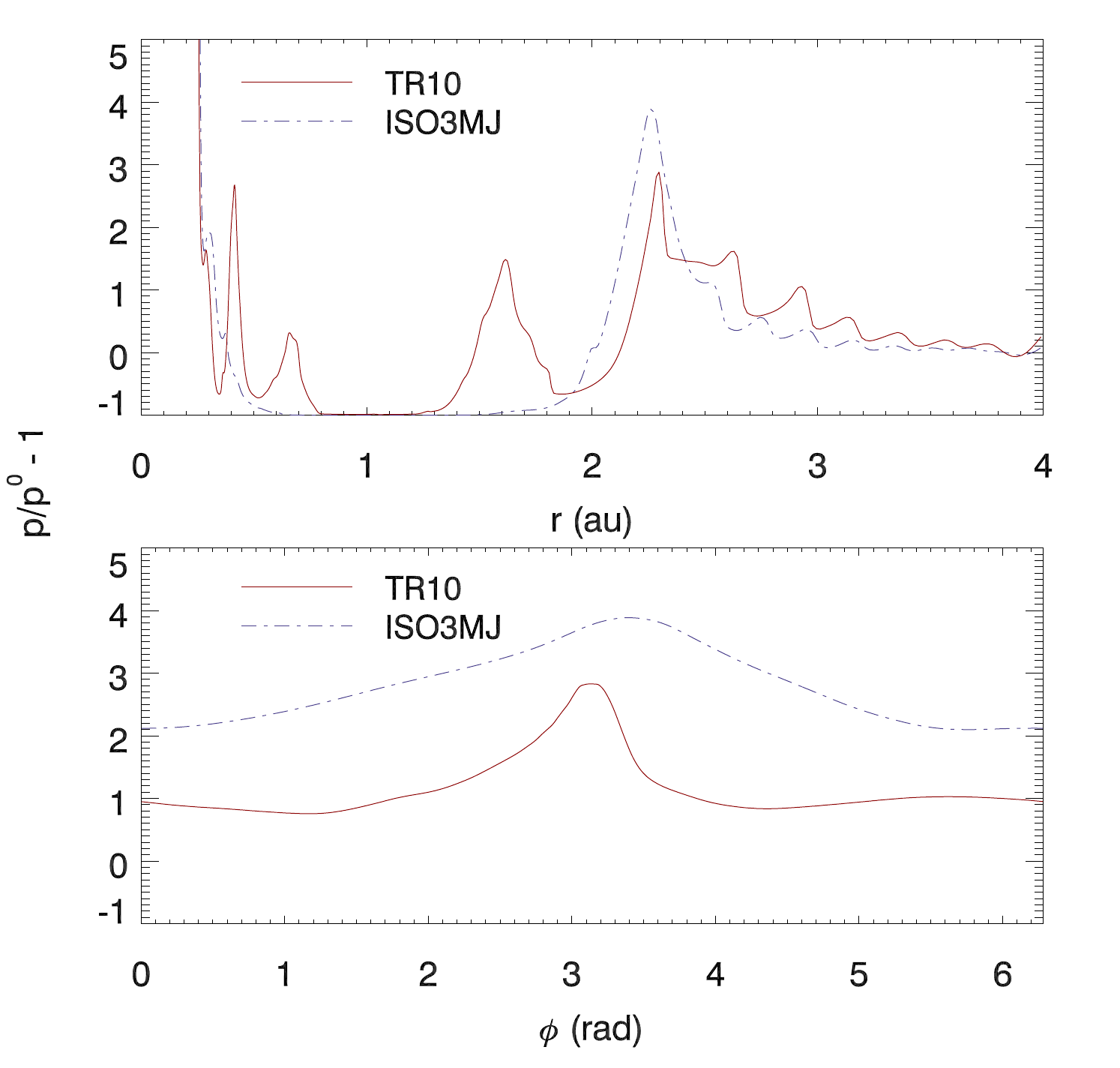}
\caption{Pressure perturbation at $t = 3500$~orbits. The top panel shows a cut of the pressure for $\phi$ equals to the secondary vortex center. The bottom
panel shows a cut of the pressure for $r$ equals to the secondary vortex center. The red solid line represents simulation TR10, whereas the slate blue
dotted-dashed line represents simulation ISO3MJ.}
\label{fig:pressure}
\end{figure}

	Vortices are able to trap dust particles, because they are local pressure bumps. The particles are attracted to the highest pressure
region, thus to the vortex center. The secondary vortex becomes extensively spread in the azimuthal direction for $\Omega\tau \geq 1.0$ and $M_p = 3M_J$. In order to make sure that
these nonaxisymmetric structures can still trap dust particles, we checked their pressure profiles. Figure \ref{fig:pressure} shows a
radial cut of the pressure for $\phi$ equals to the vortex center and an azimuthal cut of the pressure for $r$ equals to the vortex
center. We show the cases of $\Omega\tau = 10.0$ and $M_p = 3M_J$, since these are the simulations that present the most spread vortices in the
azimuthal direction. We can observe a pressure bump in both radial and azimuthal directions; however, in the azimuthal direction the bump of ISO3MJ is
very smooth. \cite{bib:birnstiel13} showed that a very smooth pressure bump in the azimuthal direction is
still sufficient to trap mm and cm particles in the vortex center. Therefore the secondary vortices should be able to trap dust particles,
leading to an asymmetric global dust distribution.

\subsection{Oph IRS 48}

	The system Oph IRS 48 is a good candidate to host a vortex-like structure induced by a planet \citep{bib:marel13}. The continuum emission ALMA
observations at $0.44$~mm revealed a high-contrast asymmetry in the disk of this system, which was interpreted as existing due to the presence of an anticyclonic
vortex \citep{bib:marel13}. Besides, this system shows a central cavity in CO line observations, which was explained as a gap opened by a massive
planet \citep{bib:brown12}. \cite{bib:marel13} ran a FARGO simulation considering the parameters of this system to get the gas density distribution. Later on the
result from the HD simulation was used as the initial condition in a dust evolution code to get the expected continuum emission. They were able to
roughly reproduce the ALMA observation; however, there is a debate regarding the location of the vortex. If the planet is located at $20$~AU, the vortex
is expected to be located at most at $\sim 45$~AU, nonetheless it is located at $\sim 63$~AU.

\begin{figure}[!htb]\centering
\includegraphics[width=.5\textwidth]{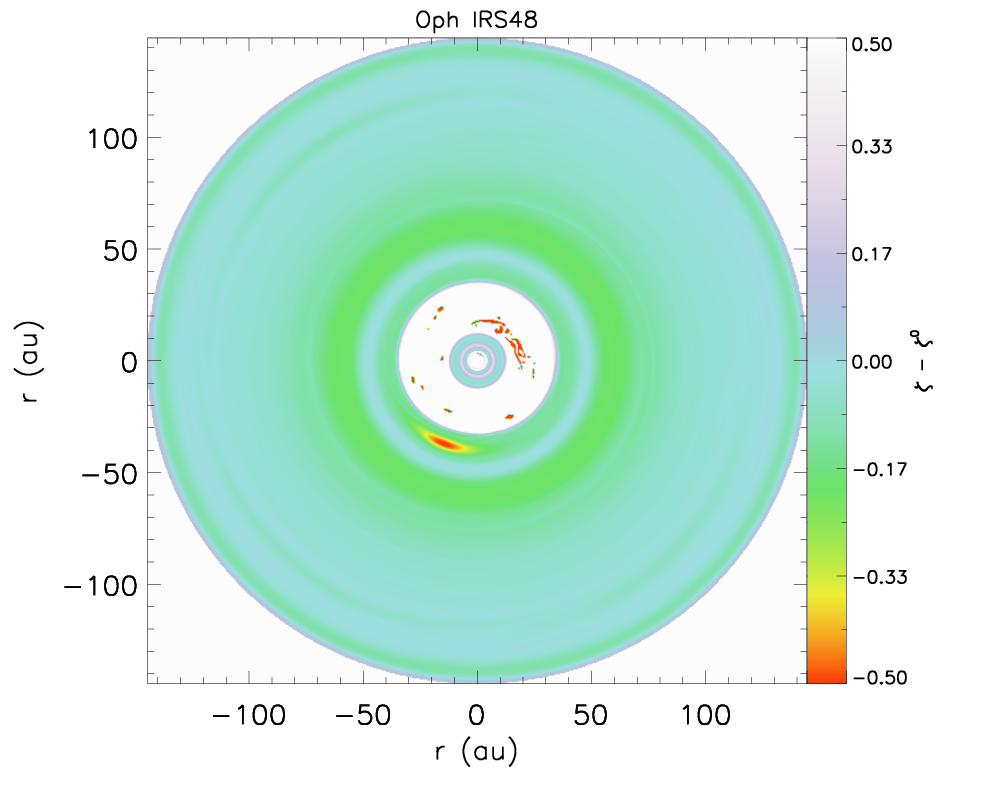}
\caption{Potential vorticity with the Keplerian profile subtracted for $t~=~700$~orbits. The color bar was truncated from $-0.5$ to $0.5$ in order
to obtain a higher contrast.}
\label{fig:ophirs48}
\end{figure}

	We run a simulation using the same setup as \cite{bib:marel13}, with the difference that here no viscosity was included, but instead we used thermal relaxation, and
initialized the disk with $H/r$ constant. We
observed vortex formation at the outer edge of the planetary gap, the vortex position is roughly $40$~AU; however, we did not observe a second generation
of vortices. The disk considered was much
larger (from $2$ to $150$~AU) than the one of our benchmark cases (from $0.25$ to $4.0$~AU), therefore the resolution may not have been
sufficient to solve the secondary vortex. Figure \ref{fig:ophirs48} shows the potential vorticity for this
simulation after $700$ planetary orbits. We can see that $\zeta$ is negative around $60$~AU, therefore a secondary vortex could be formed in that
region.

\begin{figure}[!htb]\centering
\includegraphics[width=.5\textwidth]{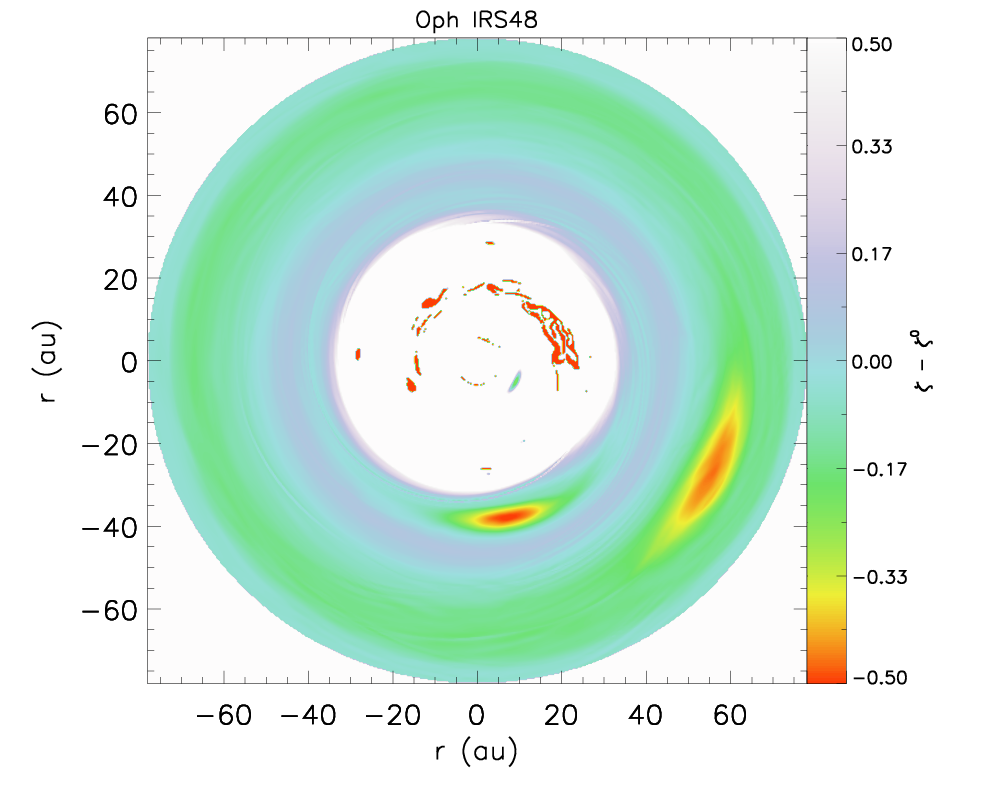}
\caption{Potential vorticity with the Keplerian profile subtracted for $t~=~700$~orbits. The color bar was truncated from $-0.5$ to $0.5$ in order
to obtain a higher contrast.}
\label{fig:ophirs48b}
\end{figure}

	The ratio between the positions of the secondary and primary vortices in our benchmark cases is about $1.5$. If
this ratio is fixed, the second generation of vortices in the Oph IRS 48 system would be located at $\sim60$~AU. In order to check whether the
secondary vortex was not observed due to a numerical resolution problem, we run a second simulation considering the same planet-disk setup as before,
but integrating over a smaller disk size. We fixed the inner and outer disk radius, in order to maintain the same ratio between the planet orbital distance
and the boundaries as for the benchmark cases. The new simulation has a disk ranging from $5$~AU to $80$~AU. Figure \ref{fig:ophirs48b} shows the
potential vorticity for the smaller disk size after $700$ planetary orbits. Here, we can see the formation of a secondary vortex, located at
$\sim62$~AU, indicating that in
the previous case the numerical resolution was indeed not sufficient. This new result is a promising explanation for the location of the Oph
IRS 48's vortex, assuming a single planet at $\sim20$~AU distance from the star. It is also important to mention that at $700$ planetary orbits, the
primary and secondary vortices are present. The benchmark cases show that the primary vortex get damped before the secondary one, thus for later times, just the secondary
vortex may be present.

\section{Vortex Lifetimes and Birth Times}\label{sec:lifetime}

	In this section we obtain the primary vortex lifetime and secondary vortex birth time as a function of the thermal relaxation timescale. For this purpose, we first had to define
when the vortex is born. Once an overdensity with negative potential vorticity arises, we declare that this overdensity is a vortex. We define as
overdensity a region that possesses an average surface density at least $20\%$ higher than the average background surface density.

	The center of the vortex was defined as the position of the maximum surface density. We determined the vortex edge by looking to the position where
the potential vorticity drops to $50\%$ below its value at the vortex center. This procedure was done in both radial and azimuthal directions. Knowing the
dimensions of the vortex, we could calculate the average surface density and potential vorticity inside the vortex.

	The procedure of finding the vortex and defining its border was done until the time that the vortex can not be defined as an overdensity
anymore. Once this criterion was reached we defined the vortex as dead. In this way, the vortex lifetime
was defined as the difference between the time it is born and the time it dies. 

\begin{figure}[!htb]\centering
\includegraphics[width=.5\textwidth]{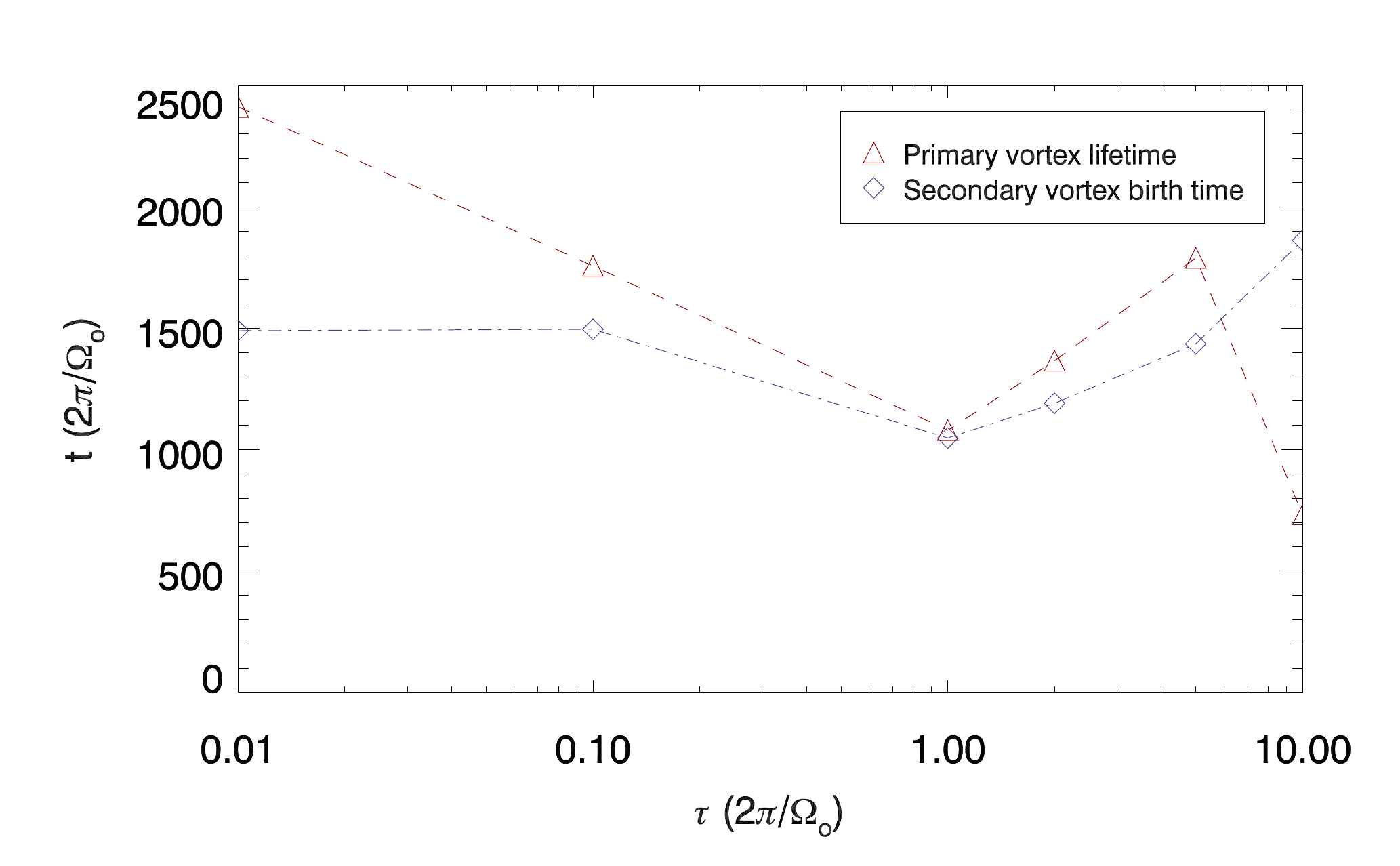}
\caption{The lifetime of the primary vortex (red dashed line) and the birth time of the secondary vortex (slate blue dotted-dashed line) as a function of the thermal
relaxation timescale.}
\label{fig:tauVStime}
\end{figure}

	Figure \ref{fig:tauVStime} presents the primary vortex lifetime as a function of the different thermal relaxation timescales. We observe a
nonmonotonic behavior that was already seen by \cite{bib:fu14} and \cite{bib:les15}. \cite{bib:les15} explained that the nonmonotonic behavior is
due to the fact that the vortex lifetime depends on (i) the decay timescale of the RWI, which decreases for increasing $\Omega\tau$, and (ii) the vortex
growth time, which increases for values of $\Omega\tau$ up to $\sim 5.0$ and then decreases for larger values. The nonmonotonic nature is a result
of this double dependence. The double peak, featuring at small $\Omega\tau$'s and $\Omega\tau = 5.0$, is due to the nonmonotonic behavior of the vortex growth
time for different $\Omega\tau$'s. Higher disk temperatures favors the RWI \citep{bib:li00,bib:lin12a}, nevertheless this effect seems to be important just
for larger $\Omega\tau$'s. The
dependence of the vortex lifetime with the vortex growth time comes to the fact that once the vortex amplitude is very large, it begins to induce shocks,
thus the vortex looses energy through shock dissipation and starts to decay. The inviscid approximation may have influenced the estimation of the
vortex lifetimes, since it is inversely dependent on the viscosity magnitude \citep{bib:valborro07,bib:ataiee13,bib:fu14}. Nonetheless, it is clear
that it did not influence the qualitatively behavior of vortex lifetimes as a function of thermal relaxation timescales, since our results are in
agreement with \cite{bib:fu14} and \cite{bib:les15}.

	We also plot in Figure \ref{fig:tauVStime} the time when the secondary vortex is born. A nonmonotonic behavior is also observed and the curves are shifted by a few hundreds of
planetary orbits, with exception for $\Omega\tau = 10.0$. Since the primary vortex is born in a scale of tens of planetary orbits, it is clear that the
secondary vortex is always born before the death of the primary vortex, again with exception for $\Omega\tau~=~10.0$. The time taken for
the secondary vortex to be born is correlated to the time that the primary
vortex needs to deplete the mass on its orbit. Therefore it depends on the vortex growth time and the accretion rate generated by the primary
vortex. This explains the inverse dependence of the secondary
vortex birth time and primary vortex growth time as a function of the thermal relaxation time scale.


\section{Summary and Conclusions}\label{sec:conclusions}

	Vortices can be formed as a product of the RWI and/or radial buoyancy (BI/SBI/CO). The RWI can be triggered in the walls of a planetary gap due to a sharp surface
density gradient. The disk is buoyantly unstable when the pressure and entropy gradients have the same sign, a thermal relaxation of the order of $\Omega\tau
\simeq 1.0$ also favors vortex amplification. We carried out global 2D-HD simulations of planet-disk interactions, using the PLUTO code. The aim was
to study the
long-term evolution of planet-induced vortices in inviscid disks and initially buoyantly unstable, considering several thermal relaxation timescales. Thermal relaxation is an important
ingredient to sustain radial buoyancy \citep{bib:petersen07a,bib:petersen07b}. It has also a strong impact on amplifying and
damping vortices.

	We found that radial buoyancy smoothen the surface density gradients in the wall of a planetary gap, which generates weaker
vortices. In this particular physical scenario, radial buoyancy operates against vortex amplification and survival. This effect is less pronounced for
the isothermal and quasi-isothermal states ($\Omega\tau << 1$), which is expected, since thermal relaxation is a required ingredient to sustain
radial buoyancy. The qualitative system evolution is similar for different thermal relaxation timescales and different planet masses. The major
difference is regarding the timescales of events (e.g., time required for vortex damping and mass transfer).

	The most interesting result from our simulations was the formation of a second generation of vortices. The primary vortex
creates an effective $\alpha$-viscosity that is large enough to induce accretion. We obtained $\alpha$-values in the range $\alpha~=~10^{-4}-10^{-2}$,
which agrees with what is obtained by the MRI \citep{bib:dzyurkevich10}. The accretion process depletes the mass in the primary vortex orbit, creating
a density enhancement outwards the vortex position. This bump is sufficient to trigger the RWI, leading to
the secondary vortex formation. This result is a promising
explanation for the location of the vortex in the Oph IRS 48 system \citep{bib:marel13}, which is located at $\sim63$~AU. Previous models predicted
that the vortex location
could be at most at $\sim45$~AU, assuming a single planet at $\sim20$~AU. Our model suggests that a second generation of vortices can be formed at
$\sim62$~AU, if a massive planet ($5M_J$) is assumed at $20$~AU. We suggest that further works should test the formation of a second generation of vortices in non-inviscid
disks and considering a proper treatment of the system's barycenter location, since these factors may influence the generation and sustenance of
vortices.

	We observed a nonmonotonic behavior for the vortex lifetime as a function of the thermal relaxation timescale. This result was already
observed by \cite{bib:fu14} and \cite{bib:les15}. The vortex lifetime depends on the decay of the RWI and the vortex growth time. The former
decreases as a function of the thermal relaxation timescale. The latter increases as a function of the thermal relaxation timescale up to
$\Omega\tau = 5.0$, decreasing for larger $\Omega\tau$'s. The nonmonotonic behavior and double peak observed for the vortex lifetime
is a result of this double dependence. The birth time of the secondary vortex also presents a nonmonotonic behavior. The appearance of the 
secondary vortex
is correlated to the time the primary vortex needs to deplete the mass on its orbit. Therefore it is linked to the primary vortex growth time and the accretion rate
generated by it. It is
important to remember that we considered an inviscid disk. Previous works have shown that the vortex lifetime is inversely dependent on the viscosity
magnitude \citep{bib:valborro07,bib:ataiee13,bib:fu14}. All the viscosity in our models is turbulence-triggered by the hydrodynamical instabilities. The inviscid approximation may have
quantitatively changed the vortices lifetimes; however, it did not change the qualitative behavior of vortices lifetime as a function of thermal
relaxation timescales.


~
~

\acknowledgements

We would like to thank fruitful comments by the anonymous referee, which certainly improved the quality of this manuscript. A. Lobo Gomes and H. Klahr
would like to thank financial support from the {\it Deutsche Forschungsgemeinschaft} (DFG), grant n. KL 1469/9-1. A. Lobo
Gomes is also greatful for the support by the International Max Planck Research School at
Heidelberg (IMPRS--HD). The simulations were performed on the THEO cluster at the {\it Rechenzentrum Garching} (RZG) of the Max Planck Society.


~
~

\end{document}